\documentclass[sigconf, nonacm]{acmart}
\usepackage{graphicx, subcaption, caption}
\usepackage{listings}
\usepackage{xcolor}
\definecolor{codegreen}{rgb}{0,0.6,0}
\definecolor{codegray}{rgb}{0.5,0.5,0.5}
\definecolor{codepurple}{rgb}{0.58,0,0.82}
\definecolor{backcolour}{rgb}{0.95,0.95,0.92}
\lstdefinestyle{mystyle}{
    backgroundcolor=\color{backcolour},   
    commentstyle=\color{codegreen},
    keywordstyle=\color{magenta},
    numberstyle=\tiny\color{codegray},
    stringstyle=\color{codepurple},
    basicstyle=\ttfamily\footnotesize,
    breakatwhitespace=false,         
    breaklines=true,                 
    captionpos=b,                    
    keepspaces=true,                 
    numbers=left,                    
    numbersep=5pt,                  
    showspaces=false,                
    showstringspaces=false,
    showtabs=false,                  
    tabsize=2
}
\begin{document}
\title{Open Data Fabric: A Decentralized Data Exchange and Transformation Protocol With Complete Reproducibility and Provenance}

%%
%% The "author" command and its associated commands are used to define the authors and their affiliations.
\author{Sergii Mikhtoniuk}
\affiliation{%
  \institution{Kamu Data Inc.}
  %\streetaddress{P.O. Box 1212}
  \city{Vancouver}
  \state{Canada}
}
\email{smikhtoniuk@kamu.dev}

\author{{\"O}zge Nilay Yal{\c{c}}{\i}n}
%\orcid{}
\affiliation{%
  \institution{University of British Columbia}
  %\streetaddress{}
  \city{Vancouver}
  \country{Canada}
}
\email{onyalcin@cs.ubc.ca}

\begin{abstract}
% Unrealized potential of data
Data is the most powerful decision-making tool at our disposal. However, despite the exponentially growing volumes of data generated in the world, putting it to effective use still presents many challenges. Relevant data seems to be never there when it is needed - it remains siloed, hard to find, hard to access, outdated, and of bad quality. As a result, governments, institutions, and businesses remain largely impaired in their ability to make data-driven decisions.
% Reproducibility crisis slows down improvements
At the same time, data science is undergoing a reproducibility crisis. The results of the vast majority of studies cannot be replicated by other researchers, and provenance often cannot be established, even for data used in medical studies that affect lives of millions. We are losing our ability to collaborate at a time when significant improvements to data are badly needed.

% Need for change
We believe that the fundamental reason lies in the modern data management processes being entirely at odds with the basic principles of collaboration and trust. Our field needs a fundamental shift of approach in how data is viewed, how it is shared and transformed. We must transition away from treating data as static, from exchanging it as anemic binary blobs, and instead focus on making multi-party data management more sustainable: such as reproducibility, verifiability, provenance, autonomy, and low latency.
In this paper, we present the Open Data Fabric, a new decentralized data exchange and transformation protocol designed from the ground up to simplify data management and enable collaboration around data on a similar scale as currently seen in open-source software.

\end{abstract}

\maketitle

\section{Introduction}
% Growing concerns about reproducibility
Modern data science and data engineering are rapidly advancing fields that feed into many other disciplines like medical and social sciences, where manipulating large datasets has long become the norm for many researchers. However, there is a growing number of concerns in those communities about the sustainability of these advancements, specifically around reproducibility and data provenance issues \cite{MiyakawaTsuyoshi2020Nrdn, baker2016reproducibility}.

% COVID highlighted issues
These issues are not new in the science community \cite{marx2013big} but became much more evident during the COVID-19 pandemic. Research efforts in the face of a global emergency required efficient collaboration around rapidly growing datasets under constant time pressure and increased attention and scrutiny \cite{Nekrutenko2020}. Under these conditions maintaining reproducibility of the results, which is currently an increasingly labor intensive process, was hardly a priority. A striking consequence of this, however, was the alarming retraction rate of the COVID-related publications from peer-reviewed journals \cite{retractRate}.

% Example: Retracted publication
One such example is the hydroxychloroquine study \cite{Mehra_Desai_Ruschitzka_Patel_2020} that was published in the influential journal, The Lancet, and claimed to use data of more than 96,000 COVID-19 patients in 671 hospitals worldwide. After the publication, the journal received a flood of concerns about the accuracy of results as some basic numbers were not checking out. The publication was later retracted when some of the included hospitals claimed that they had no arrangements to supply such data to anyone and a growing realization that provenance of source data could not be established. By that time, however, the damage was done - a data provenance issue that was left unchecked has significantly disrupted the life-saving efforts, derailed many other studies, and spread even more confusion at this critical time. The increasing awareness of these issues led to multiple action calls from the scientific community to create better collaboration platforms and manage data in a reliable way \cite{govlab, whitehouse, rdacovid19guidelines}.

% Editing data creates new datasets
We can easily identify multiple issues in modern data supply chains that contribute to these problems: publishing source data in non-machine-readable formats, data siloing, poor data quality, heterogeneous formats, poor infrastructure, lack of publisher incentives, missing feedback loop between publishers and consumers, etc. These issues make data consumers spend disproportionate amounts of time and resources on obtaining data and getting it into a usable state. The fallacy here is that since no mechanism currently exists to make such improvements reproducible and verifiable, this entire process creates yet another dataset that is entirely disjointed from its sources. This issue is present in every cycle where data is downloaded from the trusted source, modified, and then re-published. The amount of time it takes for another researcher to prove the validity of such dataset is often comparable to re-doing the whole work from scratch. Therefore, while any respectable research project starts with data from trusted data sources, it always ends up producing data that cannot be readily trusted and reused.

In this paper, we will take a detailed look at the prerequisites for building trust and collaboration and how we built the Open Data Fabric protocol to satisfy them.

\section{Design Considerations}
\begin{figure*}[h!]
\centering
\includegraphics[width=0.8\linewidth]{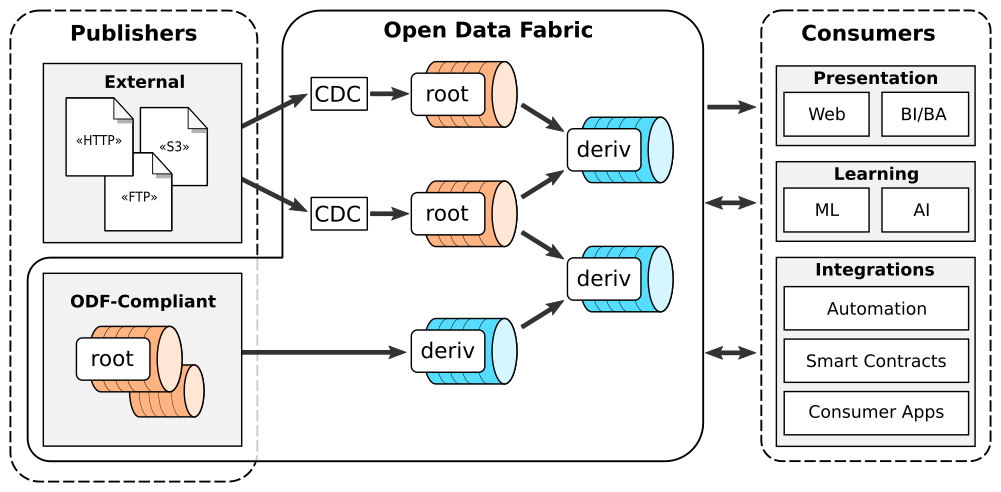}
\caption{Root and derivative dataset compose a data flow graph.}
\label{fig:dataset_kinds}
\end{figure*}

% Goals of ODF
Open Data Fabric (ODF) \cite{odf} is an open protocol specification for decentralized exchange and transformation of semi-structured data that aims to holistically address many shortcomings of the modern data management systems and workflows. Our goal is to develop a method of data exchange that would:

\begin{itemize}
\item Address the problems of reproducibility, verifiability, and provenance in modern data supply chains.
\item Create an environment of verifiable trust between participants without the need for a central authority.
\item Enable worldwide collaboration around data cleaning, enrichment, and derivation.
\item Achieve a high degree of data reuse, making quality data readily available.
\item Improve the liquidity of data by speeding up the data propagation times from publishers to consumers.
\item Close the feedback loop between data consumers and publishers, allowing them to collaborate on better data availability, recency, and design.
\end{itemize}

In this section, we will show how these goals were incorporated into the system's design.

\subsection{Data Flow}
% Data should flow freely to consumers
Raw data is rarely consumed in its original form - it often needs to go through a series of transformation, aggregation, and enrichment steps before it can be acted upon. In ODF, we make a clear distinction between the following two types of data:

\begin{itemize}
    \item \textbf{Source data}, represented in ODF by \textbf{Root Datasets}, comes directly from external systems. The organization that publishes such data has complete authority over it and is fully accountable for its veracity.
    \item \textbf{Derivative data}, represented in ODF by \textbf{Derivative Data-sets} is produced by transforming and combining other data. It is secondary to the source data but is equally important since derivative data is what's being presented to the decision-makers, used for training models, or fed into various automation.
\end{itemize}

The multi-stage process through which source data is transformed into actionable (derivative) data has many forms. It can manifest itself as reporting chains where people perform recurrent analysis and summarization tasks (e.g. payroll, company management reports), or as complex automated workflows (e.g. enterprise data pipelines). In ODF, we find it more intuitive and effective to reason about these processes as computational graphs (see Figure \ref{fig:dataset_kinds}), where data continuously flows from its sources to consumers via multiple transformation stages.

\subsection{The Foundation of Collaboration}

As shown previously, verifying the validity of derivative data can be cost-prohibitive compared to re-doing the work from scratch. This is the reason why complex multi-stage data processing chains mainly exist within the boundaries of organizations, where employees can implicitly trust one-another. Collaboration on data between the organizations today often leads to centralization - creation of data spaces where access control, audit and user privileges can be enforced. But centralization is time-consuming, expensive, and often impossible, as different parties want to maintain full control and ownership of their data.

Collaboration on data in a decentralized setting, such as research, remains unsolved. The validity of source data used in publications is often left unchecked during the review process, and the issue of trust hinges mostly on the reputation of the authors or the institution they represent. This may unfairly penalize young researchers and lesser known universities when it comes to publishing their work. 

It is therefore ODF's goal to reduce the time it takes consumers to establish validity of derivative data down to minutes and enable collaboration on data in decentralized environments where complete trust between participants is not possible.

% Drawing inspiration (not implementation) from software VCS
A modern epitome of effective collaboration between independent parties is the Open-Source Software ecosystem. What started as an exchange of ideas within a close-knit community quickly grew into a worldwide movement. The emergence of distributed version control systems (DVCS) \cite{otte2009version} ensured that the collaboration could function even at a rapidly growing scale. Many recently developed data management systems try to apply some of the ideas behind DVCS, like diff-based history and branching, directly to data \cite{dolt,deltalake}. Instead of borrowing specific technical decisions, however, in ODF we tried to understand what properties of DVCS made them so good when it comes to collaboration. We think the answer is that they help to \textit{build trust}. They can build trust even between people who have never met before, who may not even share a common language. They do so via properties like tracking the complete history of changes, allowing to attribute any change to its author, to safely roll back any previous change, and by making any malicious behavior easy to expose and remedy.

% Three pillars of collaboration
Applying these ideas to data, we have identified three pillars that would build trust and are essential to collaboration around data:

\begin{itemize}
\item \textbf{Reproducibility} - the ability to reproduce the results is the cornerstone of the scientific method without which the process and findings of one party cannot be followed by others.

\item \textbf{Verifiability} - when considering to use some data for a project it is essential to be able to understand whether it comes from a reliable publisher and whether it truthfully declares which transformations it underwent.

\item \textbf{Provenance} - regardless of how many transformation stages the data went through, it's important to be able to trace any individual value back to its source and understand what data contributed to its existence and its value.
\end{itemize}

\subsection{Reproducibility and Verifiability in Source Data}

The path towards better reproducibility and verifiability has to start at the source. The requirements here are very simple: two different parties at different times should be able to access the exact same data and validate that this data comes unaltered from the trusted source. Surprisingly, a majority of data sources today fail to satisfy these basic requirements.

% Snapshots are bad
Consider for example GeoNames \footnote{GeoNames. https://www.geonames.org/} and NaturalEarth \footnote{NaturalEarth. http://naturalearthdata.com/} - a few of the major publishers of open GIS data. On a periodic basis they provide datasets containing the latest known state of their domain. These state "snapshots" are published destructively by overwriting all the previous data. Naturally, everyone who uses the same URL to download data from these sources at different times is likely to get somewhat different data.

% Ledgers don't provide stable refs
Another example is the NYC Open Data Catalog \footnote{NYC Open Data. https://opendata.cityofnewyork.us/}, which includes some datasets with a full history of certain events. Such "ledger" only grows over time as new events get appended. This approach is much better than snapshotting as it never loses data, however, it is still quite hard for two parties to obtain the same data. Whoever downloads data first is likely to have a subset of the ledger obtained by one who downloads data later. Consumers have to rely on other means to coordinate which subset of the ledger they will be using (e.g. counting rows, using timestamps, record identifiers), which is highly unreliable (see Section 3.4).

The problems of reproducibility and verifiability of modern data start at the source data, as the bad practices like state snapshots and destructive updates are considered to be the norm for many major data publishers today. For our purposes however we have identified two crucial properties we'd like to achieve:

\begin{itemize}
\item It must be possible to obtain a stable reference to data that can be shared between parties and used to obtain the same data at any future point in time.
\item Data source has to provide a mechanism to ensure that data obtained this way was not maliciously or accidentally altered.
\end{itemize}

% Copying and versioning fallacies
One of the most prevalent strategy for achieving these properties today is to copy the entire dataset from the source onto a durable storage and assign it a version or a unique identifier. This approach is very common in the enterprise data science and popularized by data management tools like Quilt \footnote{Quilt. https://quiltdata.com/} and DVC \footnote{DVC. https://dvc.org/}. While it may work well in the closed, trusted environments such as enterprises, this approach is not suitable for a distributed setting, when working with open data, and with fast-moving data sources. Once copied, versioned snapshots essentially become fully independent datasets since no mechanism exists to reliably link them to the trusted source. Such copies contribute to the overall noise, only exacerbate the problem of provenance, and should be avoided.

% Data hubs are bad
Similar issues arise in many modern attempts to create various data hubs and data portals, such as Dataverse \footnote{Dataverse. https://dataverse.org/}, DataWorld \footnote{DataWorld. https://data.world/}, Knoema \footnote{Knoema. https://knoema.com/}, and increasingly popular data sharing on platforms like Kaggle \footnote{Kaggle. https://www.kaggle.com/datasets} and GitHub \footnote{GitHub. https://github.com/}. While the goal of simplifying discovery and federating data is noble, without a mechanism to link that data back to the trusted source all they do in fact is create more disjoint and non-trustworthy datasets.

To satisfy these properties in ODF, we changed our perspective on what "data" means to us, and make its definition more strict.

\subsection{Data Model}
\paragraph{Events}
ODF was primarily designed for mission-critical data. When data is used to gain insight and drive decision-making, discarding or modifying data is akin to rewriting history. The history of all data observed by the system must be preserved.

In the ODF, data is treated as a ledger of historical records. History only grows, it is never deleted or altered - thus all data that gets into the system is immutable. ODF treats all records in data as events or relational propositions that were believed to be true \textit{at a specific time}.
This view entirely rejects the idea of storing state snapshot data since it lacks the necessary properties. Building on the ideas of Event Sourcing \cite{EventSourcing} and Stream-Table Duality \cite{StreamTableDuality} we treat state data as a byproduct of history, which can always be reconstructed by \textit{projecting the events onto the time axis}. State data is therefore considered as a simple optimization for queries that operate with \textit{current time projections}.

\paragraph{Bitemporality}
Storing historical events alone is not enough to implement stable data references. As we mentioned previously, data consumers could agree to use a subset of the data history by utilizing some internal properties of the dataset (e.g. event timestamps, identifiers, or record counting), but it's highly error-prone.

% TODO: Shorten?
Imagine a scenario where two parties agree to use the event timestamps to delimit the data and use all events until current time T. The reproducibility in this case can be compromised by multiple factors, such as the delay it takes data to appear in the dataset, publisher performing back-fills of the events for a certain period prior to T, or corrective events with time less than T being emitted in future upon discovering errors in some old data 

To prevent such situations and issue stable references to data, ODF applies the ideas of Bitemporal Data Modelling \cite{date2002temporal,JohnstonTom2014BDTa} to the event records. We augment the schema with an extra "System Time" column, which contains the time when an event first entered the system. System Time is guaranteed to be monotonically increasing, so a stable reference to data can be as simple as having and ID of the dataset and a system time timestamp. Note that this requirement does not apply for Event Time. 

\textbf{Event time}, therefore, tells us when something happened from the outside world's perspective, and is usually the most useful one for querying and joining data. \textbf{System time}, on the other hand, gives us a reference point for when something has occurred from the perspective of the system and allows us to establish before-after relationships for data within one dataset and datasets that are part of the same computation graph.

\subsection{Reproducibility and Verifiability in Derivative Data}
For derivative data, which is obtained by transforming and combining data from other datasets, reproducibility can be thought of as having an ability to repeat all transformation steps and obtain the same results as the original. Conversely, verifiability is an ability to ensure that the data presented to you as the result of some transformation was produced without being accidentally or maliciously altered. Verifiability allows us to assess trustworthiness in two simple steps: ensuring all source data comes from reliable publishers, and auditing all applied transformations.

From these definitions, we can extract the following requirements: \textbf{determinism} - all transformations should be guaranteed to result in the same output given the same input, and \textbf{transparency} - all transformations should be known. 

% How difficult is to achieve reproducibility
These requirements are simple but extremely hard to meet in modern data science. A typical project can consist of hundreds of moving parts such as frameworks and libraries, operating systems, and hardware. Most of these components aren't purposely built with determinism in mind, so the burden of achieving reproducibility lies entirely on the person who implements the project. Achieving determinism is therefore a non-trivial problem that requires deep understanding of the execution environment and eliminating all sources of randomness. It is not surprising that such a significant undertaking is often left out completely to meet the project timelines. As a result, the modern data science is currently in a state of \textit{reproducibility crisis} \cite{MiyakawaTsuyoshi2020Nrdn, baker2016reproducibility}.

We believe that there is no way of getting around this problem - data processing systems have to be built with reproducibility in mind. Until determinism becomes an intrinsic property of such systems, we employ a series of techniques to remove as much burden as possible to ensure reproducibility (see Section 4).

% Derivative data transience
\paragraph{Derivative Data Transience} Considering that source data is immutable and all derivative transformations are deterministic, derivative data of any transformation graph in ODF can be fully reconstructed by starting from the source data and re-applying all transformations. The derived data thus can be considered as a \textit{form of caching}. As we will discuss later, this approach can potentially reduce overall data storage costs globally, since such data doesn't need to be stored durably or be heavily replicated.

\subsection{Applying Stream Processing to Historical Data}
% We were simplifying problems by throwing time away
Data science today is dominated by the batch processing workflows, whose key characteristic is treating data as a finite set of records. But a significant portion of data is not static - new data points are constantly being produced and datasets are continuously updated. The more data-driven we are, the more we will continue to the push data processing towards real-time speeds. Applying the same batch processing techniques to recent data, however, is a harmful oversimplification that exposes us to the multitude of problems associated with temporal data: data arriving late, arriving out of order, accounting for corrections that could be issued for data that has been already processed, misalignment of data arrival cadences between datasets that are being joined etc. The reliance of batch processing on data completeness often results in incorrect results, with errors concentrated in recent data - data everyone cares about the most. What's worse, as new data arrives batch workflows may produce different results for the same time periods, effectively overwriting the history - special care has to be taken to make such corrections explicit.

It would be practically impossible to write a conventional batch processing logic that correctly handles all the temporal edge cases on a mass scale. Even if we would write such logic, its complexity would negate the benefits of verifiability - what's the use of being able to audit the transformation code that is too complex to fully understand?

% Advancements in streaming
In the past few years, there has been some major advancements in the field of streaming data processing. Systems like Google Data Flow \cite{akidau2015dataflow}, Apache Spark \cite{zaharia2016apache}, Apache Flink \cite{carbone2015apache}, and Naiad and its modern implementation Timely Dataflow \cite{murray2013naiad} have developed a great apparatus for dealing with many of these problems in a very intuitive way. Stream processing paradigm acknowledges a simple fact - that data processing is a balancing act between latency and correctness. It gives user the power to control this trade-off in a fully automated way.

\begin{figure*}[htb]
\centering
\includegraphics[width=0.9\linewidth]{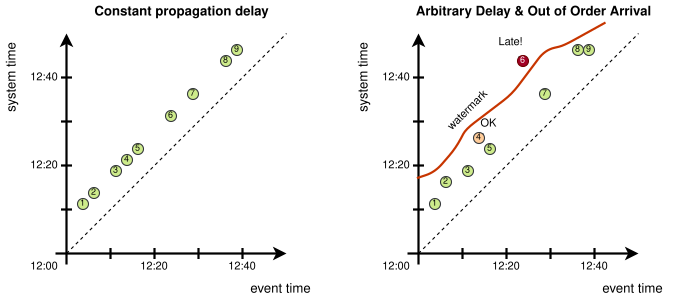}
\caption{Bitemporality and watermarks in event streams}
\label{fig:watermarks_vs_time}
\end{figure*}

One of the key mechanisms of stream processing is called the \textbf{watermark}. Watermark is a type of metadata that flows along regular data in the stream and tells the processing system that at certain system time $Ts$ with a high probability $P$ the system has observed all events prior to event time $Te$. The Figure \ref{fig:watermarks_vs_time} represents the watermark as a curve on a bitemporal event diagram that separates data that is late by an expected amount and data that is exceptionally late. Watermarks can be predictive (a system can constantly adjust it based on the observed difference between event and system time), or it can be set manually (e.g. an owner of the dataset can "hint" all data consumers that the data for the time period prior $Te$ has been fully entered). Using watermarks a stream processing system can delay the processing by the exact amount needed to handle out-of-order and late data and let users perform computations in the event time space, which is a lot more natural and easier to reason about than the arrival time. The exceptional cases of late data and backfills can also be dealt with automatically and explicitly by issuing correction or retraction events or recording that certain data points were ignored. All this makes stream processing a lot more autonomous, reliable, and composable than batch.

Conventionally, stream processing is employed to build highly responsive systems that process near real-time data. In ODF, however, we saw many benefits in applying the semantics of stream processing to historical data, even data that is updated very infrequently. Every dataset in ODF is treated as a potentially infinite stream of events. Expanding on the idea that batch processing is a special case of stream processing \cite{batchisstreaming}, we use stream processing as our primary data transformation method. 

ODF does not prescribe any specific language or framework and aims to support multiple implementations. Our first two prototype transformation engines, which we will cover in Section 4.3, use streaming dialects of SQL. An example streaming SQL query that is using stream-to-stream anti-join to detect late shipments can be seen below:

\begin{minipage}{0.95\linewidth}
\begin{lstlisting}[
language=SQL]
SELECT o.order_time, o.order_id
FROM orders as o
LEFT JOIN shipments as s
  ON o.order_id = s.order_id
  AND s.shipment_time BETWEEN
    o.order_time AND o.order_time + INTERVAL '1' WEEK
WHERE s.shipment_id = NULL
\end{lstlisting}
\end{minipage}

\vfill

% Benefits of streaming vs batch
The benefits of this approach include:
\begin{itemize}
\item Users can define a query once and potentially run it forever. This allows us to minimize the latency with which data propagates through the system.
\item Streaming queries are expressive and are closer to "which question is being asked" as opposed to "how to compute the result". They are usually much more concise than equivalent batch queries.
\item Queries can be expressed in a way that is agnostic of how and how often the new data arrives. Whether the data is ingested once a month in Gigabyte batches, in micro-batches every hour, or as a true near real-time stream - processing logic can stay the same, produce the same results, and guarantee the best propagation times possible.
\item Streaming queries are declarative, while batch processing is usually imperative. Declarative nature lets us perform static analysis of queries and automatically derive provenance in many cases without tracking any extra data.
\item High-level abstractions like windowing, watermarks, and triggers allow users to produce maximally correct results within the configurable latency window, preventing cascading error effects typically seen in similar batch workflows.
\end{itemize}

\begin{figure*}[htb]
\centering
\includegraphics[width=0.8\linewidth]{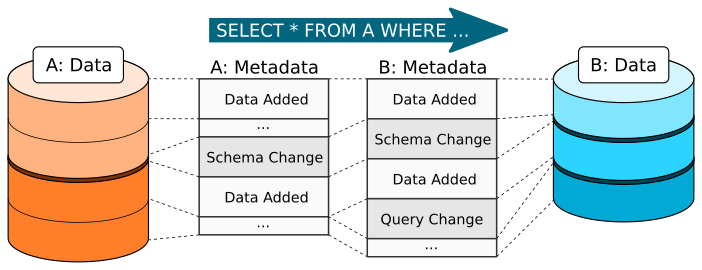}
\caption{Metadata captures all dataset life-cycle events}
\label{fig:metadata_slices}
\end{figure*}

The property of low latency that emerges from the described combination of our data model and stream processing is worth highlighting. Currently, we often see situations where critical data (e.g. employment situation reports, or COVID-19 cases reports in the early days of pandemic) is released so infrequently that it always has a dramatic effect on the stock markets, or prompts over-corrective actions from leaders and governments. We believe that the prevalence of batch workflows is a major contributing factor that adds significant delays to every data transformation stage, no matter how well-automated it is. By using stream processing and allowing people to define transformations in a way agnostic to how often data arrives the end-to-end propagation time of data can be reduced from weeks/months to mere seconds. As new data arrives, it is immediately made available to the consumer in its most usable form. This satisfies one the main guiding design principles of ODF that the frequency with which data is presented to consumers should be optimized for their experience and usability, not dictated by limitations of the data pipeline. 

\subsection{Metadata Tracking}
% Metadata as digital passport
In ODF, data never appears in the system alone, as we would not be able to tell whether it can be trusted. Metadata, therefore, is an essential part of a dataset. It contains every aspect of where the data came from, how it was transformed, and everything that ever influenced how data looks like throughout its entire lifetime (see Figure \ref{fig:metadata_slices}). We think of metadata as a digital passport of data that is instrumental to reproducibility, verifiability, and data provenance.

% Transform graph needs to be dynamic
So far, when talking about datasets, we have been assuming that the inputs, the transformations, the schema of the result and more were constant. While new events can be ingested and would propagate through the system, the processing graph itself was frozen in time. This was a deliberate oversimplification.
As the nature of businesses change, new requirements arrive, defects are detected and fixed - it's not a matter of if but when the time comes to make changes. Calling data a potentially infinite stream would not make sense without providing a way to improve and evolve it over time. Having to create a new dataset every time you need to change the schema or update the transformation would mean that the entire downstream processing graph would have to be rebuilt from scratch.
In order to support backward-compatible schema changes, evolution of transformations over time, and to provide a way to correct past mistakes in data and queries we have developed a ledger-based mechanism for recording the dataset metadata over time. We will look at it in more detail when discussing the Metadata Chain (see Section \ref{sec:metadata_chain}). 

\subsection{Provenance}
% Different from verifiability
An ability to easily understand how a specific data point came to be is crucial for building trust and showing that data can be relied upon. While verifiability can tell us which data sources were used to produce the results and which transformations were performed, this is often too coarse-grained. Provenance, on the other hand, is the ability to trace a specific piece of data back to its ultimate source, understand which events have directly contributed to its value, and what data was considered to determine its existence in the output.

% Largely unsolved
Provenance \cite{cheney2009provenance} is an is extensively studied problem in modern data science \cite{simmhan2005survey}, but in practice it still didn't fully manifest in any widely applied system or a state of the art enterprise data pipeline. Most solutions implement it only on the dataset level, as a simplified form called \textit{lineage}. Granular provenance is much challenging to achieve as it needs to span through virtually every component of the data pipeline, and potentially across many independent systems. Even if fully implemented, its value would be limited due to mutable nature of data in many modern data systems.

% Multi-level support
We believe that not being able to answer provenance questions fast can undermine the credibility of even fully trustworthy data, so ODF was designed with complete provenance in mind and supports it on multiple levels. Firstly, the metadata tracking creates a link between output and input data blocks for every iteration of a transformation. This lets ODF improve provenance granularity by limiting the search space from entire datasets to small blocks of data within them. Since both data and metadata are immutable, this link is never lost. Secondly, the declarative nature of the streaming transformations allows us to easily analyze the structure of queries. For simple map-style queries, provenance can be derived automatically without tracking any additional information. Thirdly, for more complex queries we require provenance to be supported by an underlying data processing engine \cite{provenanceSQL}. ODF defines the provenance query API that specific engine implementations need to implement (see Section \ref{sec:engine}).

\begin{figure*}[ht]
\centering
\includegraphics[width=0.9\linewidth]{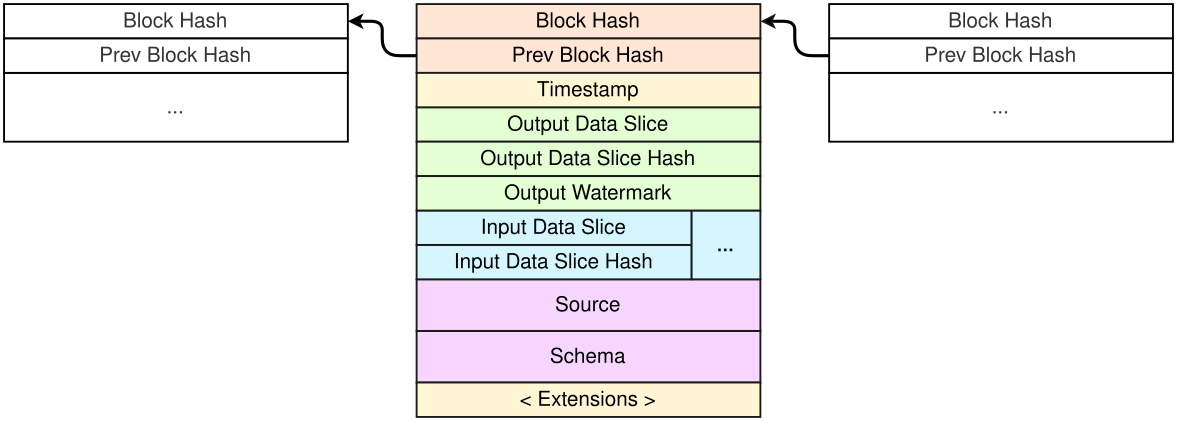}
\caption{Metadata as a chain of immutable blocks}
\label{fig:metadata_chain}
\end{figure*}

\subsection{Data Sharing}
As a distributed protocol, ODF was designed with data sharing efficiency in mind and the previously covered features directly contribute towards this goal.

The source data is irreducible by definition. ODF makes no assumptions that this data can be retrieved from anywhere else in case its lost, thus every peer that publishes root datasets is responsible for storing them durably, in a replicated and highly-available way.
The derivative data, as we covered, is considered transient; therefore all parties that publish derivative datasets can use the cheapest data hosting available (or no hosting at all) without the need for durability or heavy replication.

The immutability property of both data and metadata ensures they can be easily and safely replicated without complex synchronization mechanisms. Metadata being cryptographically linked to the raw data means that there is no need to encrypt the data itself or use a secure channel for distributing it unless we want data to remain private. It is usually sufficient to securely distribute the metadata and use it to establish the authenticity of the downloaded data. Metadata is several orders of magnitude smaller than associated data, so it can be easily hosted and widely shared.

With all these properties combined, ODF has the potential to significantly reduce data storage and distribution costs globally, compared to the widespread copy-and-version approach that often results in circulation of thousands of similar copies of a dataset taken at different points in time.

\section{Implementation}
% Prototype is available
A prototype implementation of the Open Data Fabric protocol is currently available in our Kamu CLI tool \cite{kamucli}. Data transformation can be performed using two of our stream processing engine implementations based on Apache Spark \cite{zaharia2016apache} and Apache Flink \cite{carbone2015apache}. This section will introduce some key components of the ODF and technologies used to implement them.

\subsection{Data Ingestion}
% Ideal ownership of roots
Root datasets are the points of entry for external data into the system. Our vision is that eventually all root datasets will be owned by organizations with full authority over certain data (i.e. trusted publishers) and will be provided directly in ODF-compliant format.
% Ingesting external sources
As a fallback mechanism, we also provide a way to link a root dataset to some external source (e.g. using the URL) and periodically ingest its data into the system. Such configuration duplicates the data, but it is necessary to guarantee that the properties of data are preserved and insulate the rest of the system from a wide range of current bad practices in data publishing.

% CDC on snapshots
For data sources that always preserve the entire history ODF has to do little but copy and de-duplicate the data with records that were ingested previously. For data sources that publish data in a destructive way (e.g. periodic state snapshots) ODF takes on the task of "historization" - transforming state data into event form using the Change Data Capture \cite{PetrieKevin2018SCDC,ankorion2005change} (CDC) techniques.

\subsection{Metadata Chain}
\label{sec:metadata_chain}
The Metadata Chain's purpose is to capture all information and events that influenced the way data looks right now. This includes: where the data comes from, how it was processed, its schema, and current watermark (see Figure \ref{fig:metadata_chain}).

Its design borrows heavily from the existing ledger-based systems such as blockchain \cite{zheng2017overview} and version control systems \cite{otte2009version}. Just like all data in ODF, the metadata chain is append-only and immutable. It consists of individual blocks that are cryptographically linked together and also contain hashes of the associated data slices, so the overall data structure is similar to a Merkle Tree \cite{10.10073-540-48184232}.

Metadata chain is designed for extensibility and can carry other kinds of information, such as:
\begin{itemize}
\item Extra meaning and structure of knowledge (semantics, ontology)
\item Relevant policies, terms, rules, compliance, and regulations (governance)
\item License, privacy and security concerns (stewardship)
\item Information that aids discovery
\item Interoperability data to connect ODF to other ledger-based systems
\end{itemize}

We see it as a crucial building block that will allow us to collectively push the quality of data further by standardizing and automating best data science and engineering practices.

\begin{figure*}[htb]
\centering
\includegraphics[width=0.85\linewidth]{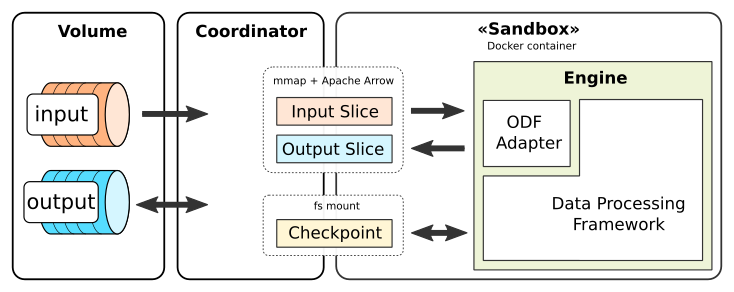}
\caption{Engine execution environment}
\label{fig:transform}
\end{figure*}

\subsection{Engine}
\label{sec:engine}
Data processing technologies are evolving rapidly. As a data exchange protocol, we would like ODF to outlive most of them and be able to adapt to new technologies as they emerge. We also want ODF to be inclusive of any data processing languages and dialects used in different branches of data science. Therefore we designed ODF to be unopinionated as to which language is used to define transformations or which framework performs them, as long as they satisfy all the criteria necessary for the protocol to function (e.g. determinism).

A specific implementation of ODF's data processing contract is called an \textbf{Engine}. Engines are responsible for applying queries defined in the datasets to input data and returning the result. For example, the two of our prototype engines based on Apache Spark \cite{zaharia2016apache} and Apache Flink \cite{carbone2015apache} allow us to transform data using a series of Streaming SQL \cite{calcite} statements. Since engines are in full control of all data transformations, they are also responsible for answering the provenance queries.

ODF takes a few extra steps to guarantee the deterministic and reproducible properties of transformations done by the engines.

\subsubsection{Execution Environment}

Engines run in a fully isolated environment called the "sandbox", implemented using the OCI Containers \cite{oci} technology (see Figure \ref{fig:transform}). Sandbox is designed to prevent engines from accessing any external resources except for the inputs and outputs of a current transformation (e.g. on the Internet or user's file system) as potential sources of undesired non-determinism.

This may sound very restrictive, and it is. After all many common data processing tasks like geolocation rely on the use of external APIs which would be inaccessible under the sandbox model. However, we strongly believe that this is a necessary step in managing data correctly. External resources like APIs are run by companies that can disappear over night, they also often evolve without following strict versioning policies - it is impossible to achieve reproducible results in such environment. Running any "black box" operations like API calls would require us to re-classify derivative datasets as source data and admit that such data is non-reproducible. Instead we envision that the software algorithms and ML models used by such transformations will be incorporated into the ODF as the engine extensions or pure data, and this transition will be one of the focus points of our future research.

\subsubsection{Engine Versioning}
To further strengthen the reproducibility guarantees of the system we associate every transformation with an exact version of an engine that was used to perform it. This excludes the possibility of any code changes in the engine producing different results than what was originally observed. For this purpose we use the full SHA digest of the engine's OCI image.

As dataset evolves over time it may start depending on too many different versions of a certain engine, unsustainably increasing the amount of images user needs to download to fully validate the dataset. We use a special \textit{engine upgrade} procedure to remedy this problem.

\subsubsection{Checkpoints \& Watermarks}
Some computations over the input data like windowed aggregations or temporal joins may require engine to maintain some state. Since engine is required to fully consume input data during transformation (as it will never see it again) some of this state may need to be preserved in between the invocations of a query. For this purpose ODF allows engines to maintain \textit{checkpoints} - a piece of opaque and fully engine-specific data used to store intermediate state. Along with data and metadata, checkpoints are an integral part of a dataset. If checkpoint is lost the entire computation will have to be restarted from scratch.

\begin{figure*}[htb!]
\centering
\includegraphics[width=0.9\linewidth]{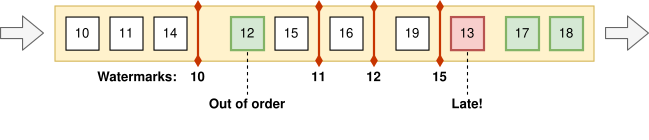}
\caption{Example use of watermarks in the event stream with an allowed lateness of t=4.}
\label{fig:watermarks_in_stream}
\end{figure*}

Every dataset in ODF also has a watermark \cite{akidau2015dataflow} - a metadata tuple $(Ts, Te)$ we described earlier. For example, if a root dataset receives new data on a monthly basis, its watermark can lag by over a month behind the wall clock time. It effectively prevents all derivative processing from proceeding past that time point until the data arrives, ensuring the correctness of computations. Figures \ref{fig:watermarks_vs_time} and \ref{fig:watermarks_in_stream} shows examples of how watermarks can prevent late processing errors in event streams. Watermarks can be set on root datasets both via fixed offset or manually, and they then fully automatically propagate through derivative datasets based on the nature of transformations. Watermarks are elevated from the checkpoints into the metadata as they are an important consumer-facing property.

\subsection{Coordinator}
The coordinator is an application responsible for maintaining the invariants and transactional semantics of the system. It handles all operations related to the metadata chain and guarantees its integrity and validity. The coordinator implements data ingestion and data sharing logic, but delegates all data processing to the engines.

Complexity of the metadata management is fully contained in the coordinator. From an engine's perspective the transformations look just like conventional stream processing, so the underlying data processing frameworks don't require any customizations or being made ODF-aware. Turning an existing data processing library into an engine is a matter of creating a thin wrapper around it that conforms to the ODF engine interface.

Besides the query execution logic, the coordinator also expects engines to implement a few extra operations related to the dataset life-cycle, such as input schema change, query change, and an engine upgrade handlers.

\section{Conclusion}
% Future - distributed network
The future of data management is a distributed network of streaming transformations, where data from trusted publishers propagates rapidly through the computational graph and is always readily available to decision-makers, automation, and AI/ML. Most of the technologies that can make this vision a reality are either already here or within our reach, but we think a mindset shift is also necessary for data-centric disciplines to abandon the local optima of ignoring temporal dimension of data, of treating data as mere binary blobs, constantly losing and rewriting our digital history.

\begin{figure}[h]
\centering
\includegraphics[width=0.9\columnwidth]{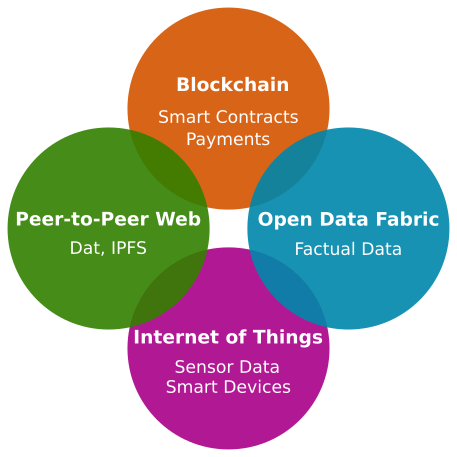}
\caption{ODF as one of the pillars of digital democracy.}
\end{figure}

% Adoption
Open Data Fabric is our attempt to achieve this state by defining the properties we want to get from data first and then designing a system around them. We saw how several key decisions like immutability of data, deterministic transformations, and using stream processing technologies coupled with ledger-based metadata tracking create a positive feedback loop with profound effects on data sharing efficiency and collaboration potential. We believe that its adoption will be a monumental step towards better data. It won't be easy, as it proposes a set of much stricter rules for processing data than what we are used to - no more manual tweaking, no more "black box" API calls - but we think the results are well worth it. Modern data processing frameworks will also have to step up to the challenge to deliver better stream and temporal processing capabilities and make determinism an intrinsic property.

% Place in web 3.0 and blockchain
In the long term, we see Open Data Fabric becoming one of the pillars of the future generation digital democracy, as the primary supply chain for structured data, which is readily consumable by the peer-to-peer web protocols and provides reliable factual data to the blockchain smart contracts to build a better more connected world. We hope others, even if they don't share our vision fully, will find some of the ideas presented here useful as we all continue to push forward the state of the art in this fascinating, remarkably complex field.

\bibliographystyle{ACM-Reference-Format}
\bibliography{biblio}


\begin{thebibliography}{32}

%%% the Web address.

\ifx \showCODEN    \undefined \def \showCODEN     #1{\unskip}     \fi
\ifx \showDOI      \undefined \def \showDOI       #1{#1}\fi
\ifx \showISBNx    \undefined \def \showISBNx     #1{\unskip}     \fi
\ifx \showISBNxiii \undefined \def \showISBNxiii  #1{\unskip}     \fi
\ifx \showISSN     \undefined \def \showISSN      #1{\unskip}     \fi
\ifx \showLCCN     \undefined \def \showLCCN      #1{\unskip}     \fi
\ifx \shownote     \undefined \def \shownote      #1{#1}          \fi
\ifx \showarticletitle \undefined \def \showarticletitle #1{#1}   \fi
\ifx \showURL      \undefined \def \showURL       {\relax}        \fi
% invisible to TeX
\providecommand\bibfield[2]{#2}
\providecommand\bibinfo[2]{#2}
\providecommand\natexlab[1]{#1}
\providecommand\showeprint[2][]{arXiv:#2}

\bibitem[\protect\citeauthoryear{Akidau, Bradshaw, Chambers, Chernyak,
  Fernández-Moctezuma, Lax, McVeety, Mills, Perry, Schmidt, and
  Whittle}{Akidau et~al\mbox{.}}{2015}]%
        {akidau2015dataflow}
\bibfield{author}{\bibinfo{person}{Tyler Akidau}, \bibinfo{person}{Robert
  Bradshaw}, \bibinfo{person}{Craig Chambers}, \bibinfo{person}{Slava
  Chernyak}, \bibinfo{person}{Rafael~J. Fernández-Moctezuma},
  \bibinfo{person}{Reuven Lax}, \bibinfo{person}{Sam McVeety},
  \bibinfo{person}{Daniel Mills}, \bibinfo{person}{Frances Perry},
  \bibinfo{person}{Eric Schmidt}, {and} \bibinfo{person}{Sam Whittle}.}
  \bibinfo{year}{2015}\natexlab{}.
\newblock \showarticletitle{The Dataflow Model: A Practical Approach to
  Balancing Correctness, Latency, and Cost in Massive-Scale, Unbounded,
  Out-of-Order Data Processing}.
\newblock \bibinfo{journal}{\emph{Proceedings of the VLDB Endowment}}
  \bibinfo{volume}{8} (\bibinfo{year}{2015}), \bibinfo{pages}{1792--1803}.
\newblock


\bibitem[\protect\citeauthoryear{Ankorion}{Ankorion}{2005}]%
        {ankorion2005change}
\bibfield{author}{\bibinfo{person}{Itamar Ankorion}.}
  \bibinfo{year}{2005}\natexlab{}.
\newblock \showarticletitle{Change data capture efficient ETL for real-time
  bi}.
\newblock \bibinfo{journal}{\emph{Information Management}}
  \bibinfo{volume}{15}, \bibinfo{number}{1} (\bibinfo{year}{2005}),
  \bibinfo{pages}{36}.
\newblock


\bibitem[\protect\citeauthoryear{Armbrust, Das, Sun, Yavuz, Zhu, Murthy,
  Torres, van Hovell, Ionescu, {\L}uszczak, et~al\mbox{.}}{Armbrust
  et~al\mbox{.}}{2020}]%
        {deltalake}
\bibfield{author}{\bibinfo{person}{Michael Armbrust},
  \bibinfo{person}{Tathagata Das}, \bibinfo{person}{Liwen Sun},
  \bibinfo{person}{Burak Yavuz}, \bibinfo{person}{Shixiong Zhu},
  \bibinfo{person}{Mukul Murthy}, \bibinfo{person}{Joseph Torres},
  \bibinfo{person}{Herman van Hovell}, \bibinfo{person}{Adrian Ionescu},
  \bibinfo{person}{Alicja {\L}uszczak}, {et~al\mbox{.}}}
  \bibinfo{year}{2020}\natexlab{}.
\newblock \showarticletitle{{Delta lake: high-performance ACID table storage
  over cloud object stores}}.
\newblock \bibinfo{journal}{\emph{Proceedings of the VLDB Endowment}}
  \bibinfo{volume}{13}, \bibinfo{number}{12} (\bibinfo{year}{2020}),
  \bibinfo{pages}{3411--3424}.
\newblock


\bibitem[\protect\citeauthoryear{Baker, Van Den~Beek, Blankenberg, Bouvier,
  Chilton, Coraor, Coppens, Eguinoa, Gladman, Gr{\"u}ning, et~al\mbox{.}}{Baker
  et~al\mbox{.}}{2020}]%
        {Nekrutenko2020}
\bibfield{author}{\bibinfo{person}{Dannon Baker}, \bibinfo{person}{Marius Van
  Den~Beek}, \bibinfo{person}{Daniel Blankenberg}, \bibinfo{person}{Dave
  Bouvier}, \bibinfo{person}{John Chilton}, \bibinfo{person}{Nate Coraor},
  \bibinfo{person}{Frederik Coppens}, \bibinfo{person}{Ignacio Eguinoa},
  \bibinfo{person}{Simon Gladman}, \bibinfo{person}{Bj{\"o}rn Gr{\"u}ning},
  {et~al\mbox{.}}} \bibinfo{year}{2020}\natexlab{}.
\newblock \showarticletitle{No more business as usual: Agile and effective
  responses to emerging pathogen threats require open data and open analytics}.
\newblock \bibinfo{journal}{\emph{PLoS pathogens}} \bibinfo{volume}{16},
  \bibinfo{number}{8} (\bibinfo{year}{2020}), \bibinfo{pages}{e1008643}.
\newblock


\bibitem[\protect\citeauthoryear{Baker}{Baker}{2016}]%
        {baker2016reproducibility}
\bibfield{author}{\bibinfo{person}{Monya Baker}.}
  \bibinfo{year}{2016}\natexlab{}.
\newblock \showarticletitle{Reproducibility crisis}.
\newblock \bibinfo{journal}{\emph{Nature}} \bibinfo{volume}{533},
  \bibinfo{number}{26} (\bibinfo{year}{2016}), \bibinfo{pages}{353--66}.
\newblock


\bibitem[\protect\citeauthoryear{Begoli, Camacho-Rodr{\'\i}guez, Hyde, Mior,
  and Lemire}{Begoli et~al\mbox{.}}{2018}]%
        {calcite}
\bibfield{author}{\bibinfo{person}{Edmon Begoli}, \bibinfo{person}{Jes{\'u}s
  Camacho-Rodr{\'\i}guez}, \bibinfo{person}{Julian Hyde},
  \bibinfo{person}{Michael~J Mior}, {and} \bibinfo{person}{Daniel Lemire}.}
  \bibinfo{year}{2018}\natexlab{}.
\newblock \showarticletitle{Apache calcite: A foundational framework for
  optimized query processing over heterogeneous data sources}. In
  \bibinfo{booktitle}{\emph{Proceedings of the 2018 International Conference on
  Management of Data}}. \bibinfo{pages}{221--230}.
\newblock


\bibitem[\protect\citeauthoryear{Carbone, Katsifodimos, Ewen, Markl, Haridi,
  and Tzoumas}{Carbone et~al\mbox{.}}{2015}]%
        {carbone2015apache}
\bibfield{author}{\bibinfo{person}{Paris Carbone}, \bibinfo{person}{Asterios
  Katsifodimos}, \bibinfo{person}{Stephan Ewen}, \bibinfo{person}{Volker
  Markl}, \bibinfo{person}{Seif Haridi}, {and} \bibinfo{person}{Kostas
  Tzoumas}.} \bibinfo{year}{2015}\natexlab{}.
\newblock \showarticletitle{Apache flink: Stream and batch processing in a
  single engine}.
\newblock \bibinfo{journal}{\emph{Bulletin of the IEEE Computer Society
  Technical Committee on Data Engineering}} \bibinfo{volume}{36},
  \bibinfo{number}{4} (\bibinfo{year}{2015}), \bibinfo{pages}{28--38}.
\newblock


\bibitem[\protect\citeauthoryear{Cheney, Chiticariu, and Tan}{Cheney
  et~al\mbox{.}}{2009}]%
        {cheney2009provenance}
\bibfield{author}{\bibinfo{person}{James Cheney}, \bibinfo{person}{Laura
  Chiticariu}, {and} \bibinfo{person}{Wang-Chiew Tan}.}
  \bibinfo{year}{2009}\natexlab{}.
\newblock \showarticletitle{Provenance in Databases: Why, How, and Where}.
\newblock \bibinfo{journal}{\emph{Found. Trends Databases}}
  \bibinfo{volume}{1}, \bibinfo{number}{4} (\bibinfo{date}{apr}
  \bibinfo{year}{2009}), \bibinfo{pages}{379–474}.
\newblock
\showISSN{1931-7883}
\urldef\tempurl%
\url{https://doi.org/10.1561/1900000006}
\showDOI{\tempurl}


\bibitem[\protect\citeauthoryear{Date, Darwen, and Lorentzos}{Date
  et~al\mbox{.}}{2002}]%
        {date2002temporal}
\bibfield{author}{\bibinfo{person}{Christopher~John Date},
  \bibinfo{person}{Hugh Darwen}, {and} \bibinfo{person}{Nikos Lorentzos}.}
  \bibinfo{year}{2002}\natexlab{}.
\newblock \bibinfo{booktitle}{\emph{Temporal data \& the relational model}}.
\newblock \bibinfo{publisher}{Elsevier}.
\newblock


\bibitem[\protect\citeauthoryear{Fowler}{Fowler}{2005}]%
        {EventSourcing}
\bibfield{author}{\bibinfo{person}{Martin Fowler}.}
  \bibinfo{year}{2005}\natexlab{}.
\newblock \bibinfo{title}{Event Sourcing}.
\newblock \bibinfo{howpublished}{martinfowler.com/eaaDev/EventSourcing.html}.
\newblock
\newblock
\shownote{Accessed: 2020-09-10.}


\bibitem[\protect\citeauthoryear{Group}{Group}{2020}]%
        {rdacovid19guidelines}
\bibfield{author}{\bibinfo{person}{RDA COVID-19~Working Group}.}
  \bibinfo{year}{2020}\natexlab{}.
\newblock \bibinfo{title}{RDA COVID-19 Recommendations and Guidelines on Data
  Sharing}.
\newblock
\newblock
\urldef\tempurl%
\url{https://doi.org/10.15497/rda00052}
\showDOI{\tempurl}


\bibitem[\protect\citeauthoryear{Johnston}{Johnston}{2014}]%
        {JohnstonTom2014BDTa}
\bibfield{author}{\bibinfo{person}{Tom Johnston}.}
  \bibinfo{year}{2014}\natexlab{}.
\newblock \bibinfo{booktitle}{\emph{Bitemporal Data: Theory and Practice}}.
\newblock \bibinfo{publisher}{Elsevier Science \& Technology},
  \bibinfo{address}{San Francisco}.
\newblock
\showISBNx{0124080677}


\bibitem[\protect\citeauthoryear{{Kamu Data Inc.}}{{Kamu Data Inc.}}{2020a}]%
        {kamucli}
\bibfield{author}{\bibinfo{person}{{Kamu Data Inc.}}}
  \bibinfo{year}{2020}\natexlab{a}.
\newblock \bibinfo{title}{{Kamu CLI - Reference implementation of ODF
  coordinator}}.
\newblock \bibinfo{howpublished}{https://github.com/kamu-data/kamu-cli}.
\newblock


\bibitem[\protect\citeauthoryear{{Kamu Data Inc.}}{{Kamu Data Inc.}}{2020b}]%
        {odf}
\bibfield{author}{\bibinfo{person}{{Kamu Data Inc.}}}
  \bibinfo{year}{2020}\natexlab{b}.
\newblock \bibinfo{title}{{Open Data Fabric - Protocol Specification}}.
\newblock \bibinfo{howpublished}{http://opendatafabric.org}.
\newblock


\bibitem[\protect\citeauthoryear{{Liquidata Inc.}}{{Liquidata Inc.}}{[n.d.]}]%
        {dolt}
\bibfield{author}{\bibinfo{person}{{Liquidata Inc.}}}
  \bibinfo{year}{[n.d.]}\natexlab{}.
\newblock \bibinfo{title}{{Dolt - Git for data}}.
\newblock \bibinfo{howpublished}{https://github.com/liquidata-inc/dolt}.
\newblock


\bibitem[\protect\citeauthoryear{Marx}{Marx}{2013}]%
        {marx2013big}
\bibfield{author}{\bibinfo{person}{Vivien Marx}.}
  \bibinfo{year}{2013}\natexlab{}.
\newblock \showarticletitle{THE BIG CHALLENGES OF BIG DATA}.
\newblock \bibinfo{journal}{\emph{Nature}} \bibinfo{volume}{498},
  \bibinfo{number}{7453} (\bibinfo{year}{2013}), \bibinfo{pages}{255}.
\newblock
\showeprint{https://www.nature.com/articles/498255a}
\urldef\tempurl%
\url{https://www.nature.com/articles/498255a}
\showURL{%
\tempurl}


\bibitem[\protect\citeauthoryear{Mehra, Desai, Ruschitzka, and Patel}{Mehra
  et~al\mbox{.}}{2020}]%
        {Mehra_Desai_Ruschitzka_Patel_2020}
\bibfield{author}{\bibinfo{person}{Mandeep~R. Mehra}, \bibinfo{person}{Sapan~S.
  Desai}, \bibinfo{person}{Frank Ruschitzka}, {and} \bibinfo{person}{Amit~N.
  Patel}.} \bibinfo{year}{2020}\natexlab{}.
\newblock \showarticletitle{RETRACTED: Hydroxychloroquine or chloroquine with
  or without a macrolide for treatment of COVID-19: a multinational registry
  analysis}.
\newblock \bibinfo{journal}{\emph{The Lancet}} \bibinfo{volume}{0},
  \bibinfo{number}{0} (\bibinfo{date}{May} \bibinfo{year}{2020}).
\newblock
\showISSN{0140-6736, 1474-547X}
\urldef\tempurl%
\url{https://doi.org/10.1016/S0140-6736(20)31180-6}
\showDOI{\tempurl}
\newblock
\shownote{Retraction in: Lancet. 2020 Jun 5;:null. Erratum in: Lancet. 2020 May
  30;: PMID: 32450107; PMCID: PMC7255293.}


\bibitem[\protect\citeauthoryear{Merkle}{Merkle}{1988}]%
        {10.10073-540-48184232}
\bibfield{author}{\bibinfo{person}{Ralph~C. Merkle}.}
  \bibinfo{year}{1988}\natexlab{}.
\newblock \showarticletitle{A Digital Signature Based on a Conventional
  Encryption Function}. In \bibinfo{booktitle}{\emph{Advances in Cryptology ---
  CRYPTO '87}}, \bibfield{editor}{\bibinfo{person}{Carl Pomerance}} (Ed.).
  \bibinfo{publisher}{Springer Berlin Heidelberg}, \bibinfo{address}{Berlin,
  Heidelberg}, \bibinfo{pages}{369--378}.
\newblock
\showISBNx{978-3-540-48184-3}


\bibitem[\protect\citeauthoryear{Miyakawa}{Miyakawa}{2020}]%
        {MiyakawaTsuyoshi2020Nrdn}
\bibfield{author}{\bibinfo{person}{Tsuyoshi Miyakawa}.}
  \bibinfo{year}{2020}\natexlab{}.
\newblock \showarticletitle{No raw data, no science: another possible source of
  the reproducibility crisis}.
\newblock \bibinfo{journal}{\emph{Molecular brain}} \bibinfo{volume}{13},
  \bibinfo{number}{1} (\bibinfo{year}{2020}), \bibinfo{pages}{24--24}.
\newblock
\showISSN{1756-6606}


\bibitem[\protect\citeauthoryear{M\"{u}ller, Dietrich, and Grust}{M\"{u}ller
  et~al\mbox{.}}{2018}]%
        {provenanceSQL}
\bibfield{author}{\bibinfo{person}{Tobias M\"{u}ller},
  \bibinfo{person}{Benjamin Dietrich}, {and} \bibinfo{person}{Torsten Grust}.}
  \bibinfo{year}{2018}\natexlab{}.
\newblock \showarticletitle{You Say 'What', i Hear 'where' and 'Why':
  (Mis-)Interpreting SQL to Derive Fine-Grained Provenance}.
\newblock \bibinfo{journal}{\emph{Proc. VLDB Endow.}} \bibinfo{volume}{11},
  \bibinfo{number}{11} (\bibinfo{date}{July} \bibinfo{year}{2018}),
  \bibinfo{pages}{1536–1549}.
\newblock
\showISSN{2150-8097}
\urldef\tempurl%
\url{https://doi.org/10.14778/3236187.3236204}
\showDOI{\tempurl}


\bibitem[\protect\citeauthoryear{Murray, McSherry, Isaacs, Isard, Barham, and
  Abadi}{Murray et~al\mbox{.}}{2013}]%
        {murray2013naiad}
\bibfield{author}{\bibinfo{person}{Derek~G Murray}, \bibinfo{person}{Frank
  McSherry}, \bibinfo{person}{Rebecca Isaacs}, \bibinfo{person}{Michael Isard},
  \bibinfo{person}{Paul Barham}, {and} \bibinfo{person}{Mart{\'\i}n Abadi}.}
  \bibinfo{year}{2013}\natexlab{}.
\newblock \showarticletitle{Naiad: a timely dataflow system}. In
  \bibinfo{booktitle}{\emph{Proceedings of the Twenty-Fourth ACM Symposium on
  Operating Systems Principles}}. \bibinfo{pages}{439--455}.
\newblock


\bibitem[\protect\citeauthoryear{Nicole Shu~Ling and Bor~Luen}{Nicole Shu~Ling
  and Bor~Luen}{2020}]%
        {retractRate}
\bibfield{author}{\bibinfo{person}{Yeo-Teh Nicole Shu~Ling} {and}
  \bibinfo{person}{Tang Bor~Luen}.} \bibinfo{year}{2020}\natexlab{}.
\newblock \showarticletitle{An alarming retraction rate for scientific
  publications on Coronavirus Disease 2019 (COVID-19)}.
\newblock \bibinfo{journal}{\emph{Accountability in Research}}
  \bibinfo{volume}{0}, \bibinfo{number}{0} (\bibinfo{year}{2020}),
  \bibinfo{pages}{1--7}.
\newblock
\urldef\tempurl%
\url{https://doi.org/10.1080/08989621.2020.1782203}
\showDOI{\tempurl}
\showeprint{https://doi.org/10.1080/08989621.2020.1782203}
\newblock
\shownote{PMID: 32573274.}


\bibitem[\protect\citeauthoryear{Noll}{Noll}{2020}]%
        {StreamTableDuality}
\bibfield{author}{\bibinfo{person}{Michael Noll}.}
  \bibinfo{year}{2020}\natexlab{}.
\newblock \bibinfo{title}{{Streams and Tables in Apache Kafka: A Primer}}.
\newblock
  \bibinfo{howpublished}{https://www.confluent.io/blog/kafka-streams-tables-part-1-event-streaming}.
\newblock


\bibitem[\protect\citeauthoryear{Otte}{Otte}{2009}]%
        {otte2009version}
\bibfield{author}{\bibinfo{person}{Stefan Otte}.}
  \bibinfo{year}{2009}\natexlab{}.
\newblock \showarticletitle{Version control systems}.
\newblock \bibinfo{journal}{\emph{Computer Systems and Telematics}}
  (\bibinfo{year}{2009}), \bibinfo{pages}{11--13}.
\newblock


\bibitem[\protect\citeauthoryear{Petrie, Potter, and Ankorion}{Petrie
  et~al\mbox{.}}{2018}]%
        {PetrieKevin2018SCDC}
\bibfield{author}{\bibinfo{person}{Kevin Petrie}, \bibinfo{person}{Dan Potter},
  {and} \bibinfo{person}{Itamar Ankorion}.} \bibinfo{year}{2018}\natexlab{}.
\newblock \bibinfo{booktitle}{\emph{Streaming Change Data Capture}
  (\bibinfo{edition}{1} ed.)}.
\newblock \bibinfo{publisher}{O'Reilly Media, Inc}.
\newblock
\showISBNx{9781492032519}


\bibitem[\protect\citeauthoryear{Simmhan, Plale, and Gannon}{Simmhan
  et~al\mbox{.}}{2005}]%
        {simmhan2005survey}
\bibfield{author}{\bibinfo{person}{Yogesh~L Simmhan}, \bibinfo{person}{Beth
  Plale}, {and} \bibinfo{person}{Dennis Gannon}.}
  \bibinfo{year}{2005}\natexlab{}.
\newblock \showarticletitle{A survey of data provenance in e-science}.
\newblock \bibinfo{journal}{\emph{ACM Sigmod Record}} \bibinfo{volume}{34},
  \bibinfo{number}{3} (\bibinfo{year}{2005}), \bibinfo{pages}{31--36}.
\newblock


\bibitem[\protect\citeauthoryear{{The Linux Foundation}}{{The Linux
  Foundation}}{[n.d.]}]%
        {oci}
\bibfield{author}{\bibinfo{person}{{The Linux Foundation}}.}
  \bibinfo{year}{[n.d.]}\natexlab{}.
\newblock \bibinfo{title}{{Open Container Initiative}}.
\newblock \bibinfo{howpublished}{https://opencontainers.org/}.
\newblock


\bibitem[\protect\citeauthoryear{{The White House}}{{The White House}}{2020}]%
        {whitehouse}
\bibfield{author}{\bibinfo{person}{{The White House}}.}
  \bibinfo{year}{2020}\natexlab{}.
\newblock \bibinfo{title}{Call to Action to the Tech Community on New Machine
  Readable COVID-19 Dataset}.
\newblock
  \bibinfo{howpublished}{https://www.whitehouse.gov/briefings-statements/call-action-tech-community-new-machine-readable-covid-19-dataset/}.
\newblock


\bibitem[\protect\citeauthoryear{Tzoumas}{Tzoumas}{2015}]%
        {batchisstreaming}
\bibfield{author}{\bibinfo{person}{Kostas Tzoumas}.}
  \bibinfo{year}{2015}\natexlab{}.
\newblock \bibinfo{title}{{Batch is a special case of streaming}}.
\newblock
  \bibinfo{howpublished}{https://www.ververica.com/blog/batch-is-a-special-case-of-streaming}.
\newblock


\bibitem[\protect\citeauthoryear{Zaharia, Xin, Wendell, Das, Armbrust, Dave,
  Meng, Rosen, Venkataraman, Franklin, et~al\mbox{.}}{Zaharia
  et~al\mbox{.}}{2016}]%
        {zaharia2016apache}
\bibfield{author}{\bibinfo{person}{Matei Zaharia}, \bibinfo{person}{Reynold~S
  Xin}, \bibinfo{person}{Patrick Wendell}, \bibinfo{person}{Tathagata Das},
  \bibinfo{person}{Michael Armbrust}, \bibinfo{person}{Ankur Dave},
  \bibinfo{person}{Xiangrui Meng}, \bibinfo{person}{Josh Rosen},
  \bibinfo{person}{Shivaram Venkataraman}, \bibinfo{person}{Michael~J
  Franklin}, {et~al\mbox{.}}} \bibinfo{year}{2016}\natexlab{}.
\newblock \showarticletitle{Apache spark: a unified engine for big data
  processing}.
\newblock \bibinfo{journal}{\emph{Commun. ACM}} \bibinfo{volume}{59},
  \bibinfo{number}{11} (\bibinfo{year}{2016}), \bibinfo{pages}{56--65}.
\newblock


\bibitem[\protect\citeauthoryear{Zahuranec}{Zahuranec}{2020}]%
        {govlab}
\bibfield{author}{\bibinfo{person}{Andrew~J. Zahuranec}.}
  \bibinfo{year}{2020}\natexlab{}.
\newblock \bibinfo{title}{Call for Action: Toward Building the Data
  Infrastructure and Ecosystem We Need to Tackle Pandemics and Other Dynamic
  Societal and Environmental Threats}.
\newblock
  \bibinfo{howpublished}{http://thegovlab.org/call-for-action-toward-building-the-data-infrastructure-and-ecosystem-we-need-to-tackle-pandemics-and-other-dynamic-societal-and-environmental-threats/}.
\newblock


\bibitem[\protect\citeauthoryear{Zheng, Xie, Dai, Chen, and Wang}{Zheng
  et~al\mbox{.}}{2017}]%
        {zheng2017overview}
\bibfield{author}{\bibinfo{person}{Zibin Zheng}, \bibinfo{person}{Shaoan Xie},
  \bibinfo{person}{Hongning Dai}, \bibinfo{person}{Xiangping Chen}, {and}
  \bibinfo{person}{Huaimin Wang}.} \bibinfo{year}{2017}\natexlab{}.
\newblock \showarticletitle{An overview of blockchain technology: Architecture,
  consensus, and future trends}. In \bibinfo{booktitle}{\emph{2017 IEEE
  international congress on big data (BigData congress)}}.
  \bibinfo{publisher}{IEEE}, \bibinfo{pages}{557--564}.
\newblock
\urldef\tempurl%
\url{https://doi.org/10.1109/BigDataCongress.2017.85}
\showDOI{\tempurl}


\end{thebibliography}

\end{document}